\documentclass[twocolumn]{aastex62}

\usepackage{graphicx}	
\usepackage{amsmath}	
\usepackage{amssymb}	
\usepackage{gensymb}

\graphicspath{{./}{figures/}}

\received{January 1, 2018}
\revised{January 7, 2018}
\accepted{\today}
\submitjournal{ApJL}

\shorttitle{Kozai-Lidov disk oscillations}
\shortauthors{Smallwood et al.}

\begin{document}

\title{Sustained Kozai-Lidov oscillations in misaligned circumstellar gas disks}

\correspondingauthor{Jeremy L. Smallwood}
\email{smallj2@unlv.nevada.edu}

\author{Jeremy L. Smallwood}
\affil{Department of Physics and Astronomy, University of Nevada \\ Las Vegas, 4505 South Maryland Parkway, Las Vegas, NV 89154, USA}

\author{Rebecca G. Martin}
\affiliation{Department of Physics and Astronomy, University of Nevada \\ Las Vegas, 4505 South Maryland Parkway, Las Vegas, NV 89154, USA}

\author{Stephen H. Lubow}
\affiliation{Space Telescope Science Institute \\ Baltimore, MD 21218, USA}



\begin{abstract}
A disk around one component of a binary star system with  sufficiently high inclination  can undergo Kozai-Lidov (KL) oscillations during which the disk inclination  and disk eccentricity are exchanged. Previous studies show that without a source of accretion, KL unstable disks exhibit damped oscillations, due to viscous dissipation, that leave the disk stable near or below the critical inclination for KL oscillations. With three-dimensional hydrodynamical simulations we show that a highly misaligned circumbinary disk that flows onto the binary components forms highly inclined circumstellar disks around each component.   We show that a continuous infall of highly inclined material allows the KL oscillations to continue. The KL disk oscillations produce  shocks and eccentricity growth in the circumstellar disks that affect the conditions for planet formation.
\end{abstract}



\section{Introduction} \label{sec:intro}

Circumbinary gas disks have a central cavity caused by the tidal torque produced by Lindblad resonances due to the central binary \citep{Artymowicz1994,Lubow2015,Miranda2015}. The cavity size is determined by the radius where
tidal torques balance viscous torques. This balance between viscous and tidal torques is determined in a one-dimensional azimuthally averaged sense and neglects nonaxisymmetric imbalances that could arise in multiple dimensions. The central binary produces an outward torque that, in the simplest models, prevents circumbinary material from flowing onto the central binary \citep{LP1974, Pringle1991}.

In spite of this central
cavity and outward torque, material can flow through the cavity as gas streams \citep{Artymowicz1996, Gunther2002, Shi2012, Dorazio2013, Farris2014, Munoz2019}.
This flow forms and/or replenishes circumstellar disks of gas and dust around each component of a binary.
For typically warm and viscous protostellar disks,
the accretion rate onto the central binary is
comparable to the rate that would occur if the
the binary were replaced by a single star, even though
viscous and tidal torques are balanced in the one-dimensional (radial) sense
at the inner edge of the circumbinary disk.  Based on simulations at a few different disc aspect ratios  $H/R$, \cite{Artymowicz1996}  originally suggested that  
 $H/R \gtrsim 0.05$,  if the turbulent viscosity parameter $\alpha$ is greater than 0.01. is required for efficient  (unimpeded) flow into the cavity. Using more accurate simulations and more $H/R$ values, \cite{Ragusa2016} suggested that $H/R  \gtrsim 0.1$ is required for efficient flow.

Circumbinary disks are sometimes observed to be misaligned to the binary orbital plane, particularly during early stages of stellar evolution. 
The pre-main sequence binary KH 15D has a low inclination precessing circumbinary ring or disk \citep{Chiang2004, Lodato2013, Smallwood2019}. The circumbinary disk around the young binary IRS 43 has an observed misalignment of $\sim 60\degree$ \citep{Brinch2016}.  There is an observed misalignment of $\sim 70\degree$ between the circumbinary disk and the circumprimary disk in HD 142527 \citep{Marino2015,Owen2017}. The $6$--$10\, \rm Gyr$ old binary system, 99 Herculis, has a nearly polar ($\sim 87\degree$) debris ring \citep{Kennedy2012,Smallwood2020}.  The gaseous circumbinary disk around HD 98800BaBb is found to be nearly polar \citep{Kennedy2019}. There is evidence that the circumbinary disk around GG Tau is misaligned
by $\sim 30\degree$ \citep{Aly2018}.



Kozai-Lidov (KL) oscillations were first studied for test particles that reside in orbits that are highly misaligned to the orbital plane of a circular orbit binary \citep{Kozai1962,Lidov1962}. The particle orbits undergo oscillations in which eccentricity and inclination are exchanged.  An initially highly inclined circular orbit can periodically obtain
high eccentricity during phases of lower inclination.
For eccentric orbit binaries, stronger effects from KL oscillations have been found to occur \citep{
Lithwick2011,
Teyssandier2013,
Liu2015}. 
Tidal dissipational effects on KL planets were analyzed by \cite{Fabrycky2007}.

A misaligned circumbinary disk may form misaligned circumstellar disks around each binary component \cite[e.g.,][]{Nixon2013}. If a circumstellar disk around a binary component is  sufficiently misaligned  with respect to the binary orbital plane and has a sufficiently low disk aspect ratio, the disk may also undergo KL oscillations in which the disk eccentricity and inclination are exchanged \citep{Martinetal2014}.  Simulations suggest that the critical inclination above which KL oscillations occurs is approximately equal to the value for particles of $\ 40\degree$, although linear theory suggests that it varies with the disc aspect ratio  \citep{Zanazzi2017, LubowOgilvie2017}. Linear theory also suggests that $H/R \stackrel{<}{\sim} 0.15$ is typically required for an equal mass binary.
  
 The properties of KL oscillations of particles and disks differ.  KL oscillations of a set of particle orbits have a tilt frequency that varies with distance from the central star, resulting in an incoherent overall tilt. A gaseous disk however can undergo a coherent tilt oscillations whose frequency is independent of radius due to the effects of pressure forces. In addition, the KL disk oscillations damp over time due to dissipation, which does not occur for test particles. 

In this Letter, we investigate the evolution of a highly misaligned KL unstable circumstellar disk that forms through accretion from a misaligned circumbinary disk.  Previous models of disks undergoing the KL mechanism found that the KL cycles damp due to viscous dissipation.  The lack of accretion of misaligned material onto the circumstellar disks in these models caused their inclinations to evolve to near or below the KL critical angle for test particles (about $39\degree$) and the cycles ended \citep{Martinetal2014,Fu2015,Fu2015b,Franchini2019}. 

However, since our simulation includes a misaligned circumbinary disk, the KL unstable circumstellar disks have a reservoir of misaligned material that continues to flow through the binary cavity  and accrete onto the circumstellar disks. As a result, we find the KL mechanism can be long-lived.  This has implications for the  accretion variability in time. The mechanism also has implications for planet formation within binary star systems.  A circumstellar disk undergoing KL oscillations experiences periodic changes in eccentricity that  affects the interaction of the gas with the solids. In this paper, we focus only on the properties of the gas. The disk can experience strong shocks during high eccentricity phases  that can affect the conditions for planet formation.
In particular, such a disk may undergo gravitational instability and fragmentation  \citep{Fu2017}. 

The outline of this Letter is as follows.
In Section~2, we describe the setup and results of our hydrodynamical disk simulation. We discuss the relevance and implications of this work in Section~3. Finally, we give concluding remarks in Section~4.

\section{Hydrodynamical Disk Simulations}


We model the evolution of an initially flat but misaligned hydrodynamical circumbinary disk with {\sc phantom} \citep{Price2018}. The {\sc phantom} code is a three-dimensional  Smoothed Particle Hydrodynamics (SPH) code that has been tested extensively for modelling misaligned accretion disks \cite[e.g.,][]{Nixonetal2012a,Nixon2013,Nixon2015,Dougan2015,Facchini2018,Smallwood2019,Aly2020}.

\subsection{Setup}

In this section, we describe the setup of the hydrodynamical simulations. 
We consider an equal mass binary with eccentricity $e_{\rm b} = 0.1$.
The binary begins at apastron at $t = 0$.  
We apply a mass sink about each star that has an accretion radius.
When a particle penetrates the accretion radius, it is accreted onto the sink and the particle's mass and angular momentum are added to the star  \cite[e.g.,][]{Bate1995}.
Larger accretion radii eliminate the need to resolve particle orbits close to the stars, which speeds up computational time significantly. However, since we want to resolve the formation of circumstellar material, the accretion radius for each star is chosen to be $0.05a$, where $a$ is binary semi-major axis.  The accretion radius is a hard boundary, meaning that all of the particles that enter are accreted and their mass and angular momentum are added to the sink particle.

 The circumbinary disk is initially misaligned by $60\degree$ to the binary orbital plane that is initially in the $x-y$ plane. The binary begins at apastron and along the $x$-axis. The disk is initially comprised of $1.5\times 10^6$ equal-mass Lagrangian particles. 
 The simulation end time is set to $45 \, \rm P_{orb}$ which is sufficient to reach a quasi-steady state,   where $\rm P_{orb}$ is the  binary orbital period.
 We simulate a low-mass circumbinary disk such that $M_{\rm CBD} = 10^{-3} M$, where $M$ is the binary mass. The particles are initially located 
in a flat circumbinary disk with an inner disk radius of $R_{\rm in} = 1.6a$ and an outer disk radius of $R_{\rm out} = 2.6a$. The inner boundary of the disk is chosen to be smaller than where  the binary tidal torque truncates the inner edge of an aligned disk \citep{Artymowicz1994}. The tidal torque for a misaligned disk is weaker which allows the inner disk radius to be closer to the binary \cite[e.g.,][]{Lubow2015,Miranda2015}.

The physical disk viscosity is modeled by using the artificial viscosity $\alpha^{\rm av}$, which is implemented in {\sc phantom} \citep{Lodato2010}. The circumbinary disk has an initial surface density profile of the form $\Sigma \propto R^{-3/2}$, and the disk aspect ratio $H/R$ is set to be $0.1$ at $R_{\rm in}$. With this prescription, the shell-averaged smoothing length per scale height $\langle h \rangle / H$ and the disk viscosity parameter $\alpha$ are constant over the radial extent of the disk \citep{Lodato2007}. We take a relatively high \cite{Shakura1973} $\alpha_{\rm SS}$ parameter of $0.1$ in order to increase resolution of the circumstellar disks by having a relatively high inflow rate from the circumbinary disk into the binary cavity. The circumbinary disk is initially resolved with $\langle h \rangle / H = 0.11$.

Traditionally, to model the evolution of a circumbinary disk, it is efficient to use a locally isothermal temperature profile centered on the binary center of mass. 
This model does not accurately apply to circumstellar disks.
Since we are interested in forming circumstellar disks from the inflow of circumbinary gas, we adopt the locally isothermal equation of state of \cite{Farris2014} and set the sound speed $c_{\rm s}$ to be
\begin{equation}
    c_{\rm s} =  {\cal{F}} c_{\rm s0}\bigg( \frac{a}{M_1 + M_2}\bigg)^q \bigg( \frac{M_1}{R_1} + \frac{M_2}{R_2}\bigg)^q,
    \label{eq::EOS}
\end{equation}
where $R_1$ and $R_2$ are the radial distances from the primary and secondary stars, respectively, and
 $c_{\rm s0}$ is a constant with dimensions of velocity. 
 Parameter $q$ is set to 3/4.
${\cal{F}}$ is a dimensionless  function of position that we define below.
 The binary total mass is $M = M_1 + M_2$, where $M_1$ and $M_2$ are the masses of the primary and secondary stars, respectively. This prescription of the sound speed ensures that the temperatures in the circumprimary and circumsecondary disks are mainly set by the primary and secondary stars, respectively.  For $R_1, R_2 >> a$, the sound speed is set by the distance from the binary centre of mass. 

To further improve the resolution of the circumstellar disks, we introduce function $\cal{F}$ in Equation (\ref{eq::EOS}) to decrease the sound speed close to each binary component, thus leading to a longer viscous timescale and more mass in the steady state disk. We take
\begin{equation}
  \cal{F}=\begin{cases}
   \sqrt{0.001}, & \text{if $R_1\, {\rm or}\, R_2 < R_{\rm c}$},\\
    1, & \text{otherwise},\\
  \end{cases}
\end{equation}
where $R_{\rm c}$ is the cutoff radius. We ran extensive tests to determine the optimal value of $R_{\rm c}$. 
 If this value is too small (close to the binary), disk formation will be suppressed, due to the low resolution caused by the larger separation between the particles. 
On the other hand, a cutoff radius that is too far from the binary components will affect the circumbinary disk.
We adopt a cutoff radius of $R_{\rm c}=0.35a$ from each binary component.  

Beyond this cutoff distance from  a binary component, the gas flow through the gap has not encountered the bound material orbiting that binary component. Consequently, there are no shocks beyond the cutoff radius involving that component's circumstellar disk.
 With this prescription, the disk aspect ratio of the circumstellar disks at radius $r=0.1a$ is $H/R \sim 0.01$. This aspect ratio is one-tenth of the disk aspect ratio at the initial inner circumbinary disk radius. 
Since the motion of the gas streams that flow through the gap is  supersonic, the gas temperature has little effect on that flow.
The properties of the gas flow that transports material from the circumbinary disk towards the circumstellar disks should then be similar with and without this factor. 
 The factor $\cal{F}$ results in a stronger shock when the inflowing gas encounters a circumstellar disk
that in turn results in a higher post-shock density and hence higher resolution.
This factor also significantly increases the resolution within the forming circumstellar disk, however, the resolution is not as high as that of the circumbinary disk. We have tried numerous additional approaches to increase the resolution of the circumstellar disks while also still resolving the circumbinary disk but to no avail. The current approach presented in this Letter is the best option we have found  for the  smoothed particle hydrodynamics simulations we present.


\subsection{Results}
In this section, we explore the formation of circumstellar disks from material initially originating from a highly misaligned circumbinary disk. 
We extract the disk parameters as a function of time. To analyze the data on the circumbinary disk, we average over all particles which range from the innermost bound circumbinary particle out to a distance of  $3a$. 
For the circumstellar disks, we average over all particles that are bound to each binary component (i.e., the specific energies, kinetic plus potential, of the particles are negative, neglecting the thermal energy). For each disk, we calculate the mean properties of the particles such as the surface density, inclination, longitude of ascending node, and eccentricity.

\begin{figure} 
\includegraphics[width=\columnwidth]{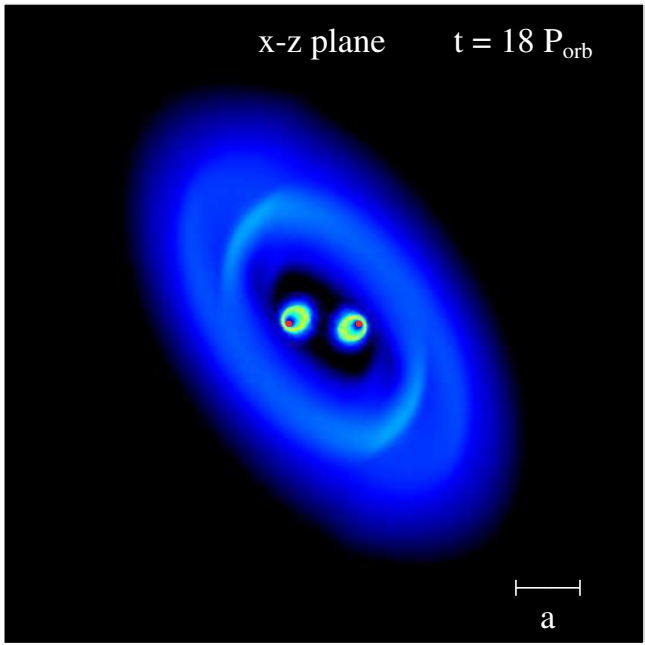}
\centering
\caption{The formation of circumstellar disks from a low-mass circumbinary disk that is initially inclined by $60\degree$ at $t = 18\, \rm P_{orb}$. The eccentric orbit binary components are denoted by the red circles.  The color denotes the gas density using a density weighted interpolation, which gives a mass-weighted line of sight average. The yellow regions are about three orders of magnitude larger than the blue. At this time, the circumstellar disks are gaining eccentricity due to the KL mechanism. The view is in the $x$--$z$ plane. }
\label{fig::splash}
\end{figure}

\begin{figure} 
\includegraphics[width=\columnwidth]{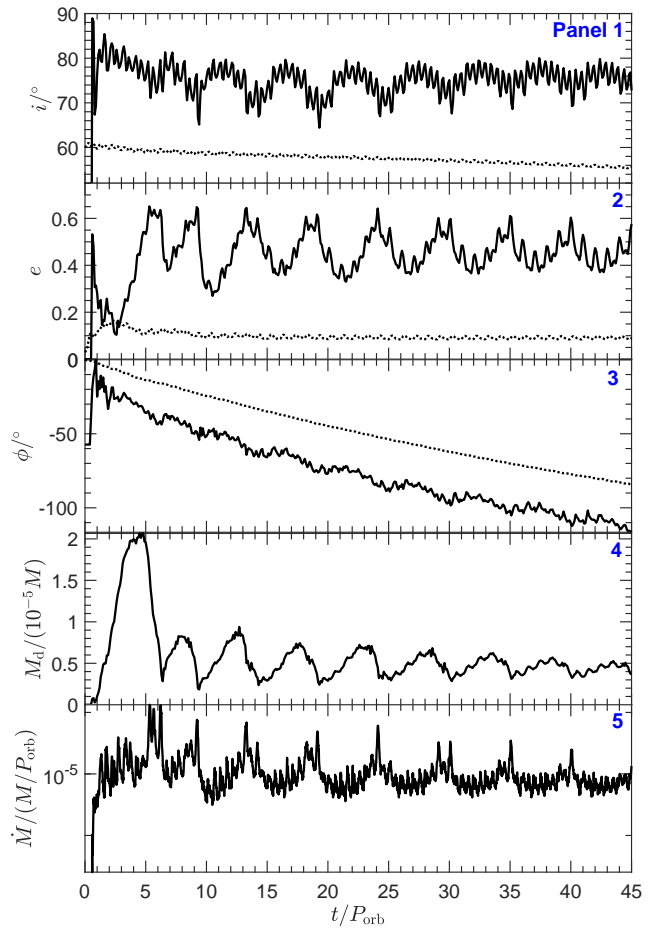}
\centering
\caption{Disk parameters for the  circumprimary disk (solid) and circumbinary disk (dotted) as a function of time in units of the binary orbital period. The disk parameters are inclination $i$ (Panel 1), eccentricity $e$ (2), longitude of the ascending node $\phi$ (3), disk mass $M_{\rm disk}$ (4), and mass accretion rate $\dot{M}$ (5). The circumbinary disk results are only shown for the top three panels.
}
\label{fig::disc_params}
\end{figure}

Figure~\ref{fig::splash} shows the formation of circumstellar disks from the highly misaligned circumbinary disk at a time $t = 18\, \rm P_{orb}$. Both the circumprimary and circumsecondary disks are becoming moderately eccentric due to the KL mechanism. Figure~\ref{fig::disc_params} shows the inclination, eccentricity, longitude of the ascending node, disk mass, and accretion rate as a function of time in units of the binary orbital period. We plot the properties of only the circumprimary and circumbinary disks. We note the circumsecondary disk has similar disk properties to the circumprimary disk. Panel 1 shows disk inclination evolution. The circumprimary disk initially forms with an inclination of $\sim 80\degree$. Note that the initial misalignment of the circumbinary disk is $60\degree$ and remains at a similar inclination throughout the simulation. Therefore, the circumstellar disks form at a higher inclination  than the initial circumbinary disk inclination.  The circumprimary and circumsecondary disks undergo KL cycles, since the inclinations are above the  critical angle for test particles of about $39\degree$. For a gas disk, the criteria to induce KL cycles also depends on disk aspect ratio \citep{Zanazzi2017, LubowOgilvie2017}. If the aspect ration $H/R$ is too high, greater than roughly 0.15 for an equal mass binary, the KL oscillations can be suppressed.

Panel 2 shows the disk eccentricity as a function of time.  After $\sim 2\, \rm P_{orb}$, the material that falls in has eccentricity of around 0.2, before any KL oscillations are established. At later times there are  repeated oscillations of eccentricity growth. The maximum eccentricity induced is  $\sim 0.63$. Comparing the top two panels in Fig.~\ref{fig::disc_params}, we see that as the eccentricity increases, the inclination decreases, and vice-versa, which is indicative of the KL mechanism operating. The period of the KL oscillations is about $6 P_{\rm orb}$ that is similar to the period found in previous work on KL disks \citep{Martin2014}.
The KL cycles that are induced do not quickly damp as previous models have found due to the fact the circumstellar disks are continuously fed by high inclination material originating from the circumbinary disk.  However, the amplitude of the oscillations decrease somewhat with time, which is due to the depletion of mass in the circumbinary disc. At $45\, \rm P_{orb}$, the circumbinary disc has lost $\sim 40\%$ of its initial mass. Moreover, the minimum eccentricity during the KL oscillations is larger than the eccentricity of the infalling material of $\sim 0.2$. This means that the KL unstable disks are able to maintain a disk structure, rather than becoming completely accreted during each high eccentricity phase and then recreated.

Panel 3 in Fig.~\ref{fig::disc_params} shows the longitude of the ascending node as a function of time. The circumstellar disk begins to precess as it is formed and it precesses at a faster rate than the circumbinary disk. 
Panel 4 shows the evolution of the circumprimary disk mass.  Due to the large disk aspect ratio and high viscosity, 
there is a high infall rate of material into the cavity within the first $5\, \rm P_{orb}$. This causes the circumstellar disk mass to grow  rapidly. However, beyond $5\, \rm P_{orb}$, the mass in both circumstellar disks reach a quasi-steady state. 
The accretion rate peaks at times when the eccentricity is locally maximum. 
At the times of each maximum (minimum) in eccentricity, the disk's mass decreases (increases) due to increased (decreased) accretion rate onto the binary components. Panel 5 shows the accretion rate on a log scale as a function of time. 

\begin{figure} 
\includegraphics[width=\columnwidth]{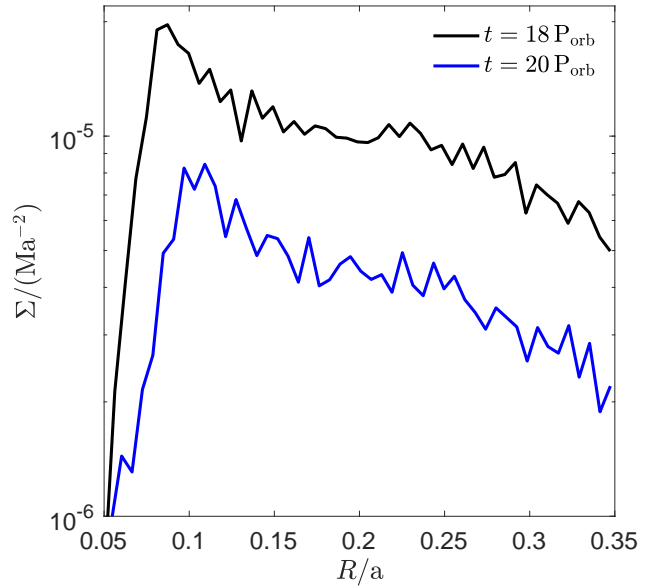}
\centering
\caption{The surface density profile as a function of radius for an inclined circumprimary disk. The surface density profile is shown at two different times: $18\, \rm P_{orb}$ (black) and $20\, \rm P_{orb}$ (blue).}
\label{fig::disc_size}
\end{figure}

Lastly, we examine the evolution of the surface density of the circumprimary disk. Since we are modeling an equal mass binary, the evolution of the circumprimary and circumsecondary disks are similar. Figure~\ref{fig::disc_size} shows the circumprimary disk surface density at times $t = 18 \, \rm P_{orb}$  and $20\, \rm P_{orb}$. The disk forms at a time $\sim 2\, \rm P_{orb}$. 
 At $t = 18 \, \rm P_{orb}$, there is a locally maximum disk mass and an increasing eccentricity. At a time $t = 20 \, \rm P_{orb}$, there is a locally minimum disk mass and decreasing eccentricity. 
Since we have found that the KL cycles are continuous, the surface density of the circumstellar disks  continuously oscillates between the black and blue lines.

\section{Discussion}

Binary stars are common within the Galaxy \cite[e.g.,][]{Duquennoy1991,Raghavan2010}. It has been estimated that roughly 50\% of the discovered exoplanets may be hosted by binary systems \citep{Horchetal2014, Deacon2016,Ziegler2018}. For example, the binary system $\gamma$ Cep AB has a separation of $\sim 20\, \rm au$ and hosts a giant planet around the primary star, $\gamma$ Cep Ab \citep{Hatzes2003}. 
The planet formation process is thought to take place in a protoplanetary disk. For the cases of binary star systems, material from a circumbinary disk can feed the circumstellar disks in which planets may form. The effects of KL oscillations of the circumstellar disks found in this Letter play an important role in affecting the formation of planets around each individual binary component involving a misaligned circumbinary disk. 

The KL mechanism operates in binary systems in cases in which the circumstellar disks are highly misaligned to the binary orbital plane. \cite{Fu2017} found that strong shocks are generated in the circumstellar disks when the eccentricity is high. Their models had KL cycles that damped due to viscous dissipation. Since we have shown that KL cycles can be sustained for longer periods of time, shocks will be generated on each KL cycle, which will impact the formation of S-type planets in binaries.  In sufficiently massive discs shocks may trigger the gravitational instability (GI), which is a possible theory to form giant planets \citep{Boss1997}. Therefore, the KL mechanism can cause disk GI, which leads to fragmentation  and efficient formation of giant planets \cite[][]{Fu2017}. 

 Massive circumbinary discs will provide both an accretional torque and gravitational torque on the binary, which will affect the binary orbital evolution \citep{Munoz2019}, and thus affect the misalignment evolution \citep{MartinLubow2019}. For an initially low eccentricity binary, $e_{\rm b} = 0.1$ (which is used in our hydrodynamical simulations), \cite{Artymowicz1991} found that the binary eccentricity is increasing due to gravitational effects of the disc.  For fixed binary semi-major axis, circumstellar disk radii decrease with increasing binary eccentricity \citep{Artymowicz1994, Miranda2015}. 

The presence of a circumbinary disk allows for a reservoir of material that accretes onto the forming circumstellar disks. Throughout the simulation, the misalignment of the circumbinary disk remains nearly constant. However, the resulting circumstellar disks form initially at a higher inclination and then proceed to undergo the KL mechanism where the inclination oscillates. The KL cycles are more periodic when including accretion of material from the circumbinary disk. However, the cycles do not repeat as precisely as in the case of test particle orbits. This is a result of the KL mechanism being  sensitive to the binary and disk parameters that are changing in time \cite[e.g.,][]{Fu2015}.  For example, the circumbinary disk depletes to about 
40\% the initial disk mass at the end of the simulation. During this process, material is continuously being transported with high inclination to the lower inclination circumstellar disk. Accretion keeps the inclination of the disk high and allows the KL cycles to continue. However, if circumstellar disk material were to accrete on a timescale shorter than the KL oscillation period, we would not expect the KL oscillations to operate.  In this work, we consider an equal mass binary. However, for an unequal mass binary, we expect that the KL oscillations would be qualitatively similar. But the gas stream feeding will not be the same for the two circumstellar discs
and their oscillation properties will differ.

Previous studies have suggested that tidally induced shocks could play an important role in causing disk accretion
\citep[e.g.,][]{Goodman2001, Ju2017}. From Panel 5 in Figure~\ref{fig::disc_params} and also Figure~\ref{fig::disc_size}, we see that there are  KL induced variations of mass accretion rate and disk density that are likely
attributable to accretion by shocks. Some of the accretion may also be due to higher eccentricity material entering the mass sink.

\section{Conclusions}

In this Letter, we have shown that the KL mechanism can be repeatedly sustained without damping when the circumstellar disks are replenished from a highly misaligned circumbinary disk. For example, a circumbinary disk with an initial misalignment of $60\degree$ to the binary orbital plane forms circumstellar disks that are initially inclined by $\sim 80 \degree$. Since the inclinations of the newly formed disks are larger than the KL critical angle and the disk aspect ratio criterion is met, the disks undergo KL oscillations, where the inclination is exchanged for eccentricity and vice versa. Previous models of the KL disks are damped due to viscous dissipation in the absence of accretion from a circumbinary disk. By modelling the formation of circumstellar disks from the circumbinary disk, the KL cycles are long-lived due to a large reservoir of noncoplanar material infalling from the circumbinary disk and continuously accreting onto the circumstellar disks.  The longevity of the KL mechanism has important implications for planet formation in binary star systems.

\section{Acknowledgements}
We thank the anonymous referee for the helpful feedback which improved the quality of the manuscript. We thank Daniel Price for providing the {\sc phantom} code for SPH simulations and acknowledge the use of SPLASH \citep{Price2007} for the rendering of the figures. Computer support was provided by UNLV's National Supercomputing Center. We acknowledge support from NASA XRP grant 80NSSC19K0443.





\bibliography{ref}



\end{document}